\documentclass[12pt,a4paper]{article}
\usepackage{epsfig}
\usepackage{graphicx}                  
\usepackage{ae}
\usepackage{amsmath}
\usepackage{amssymb}
\usepackage{graphics}
\usepackage{dcolumn}
\usepackage{bm}
\begin{document}
\begin{center}

{\bf \Large Studies of the effect of rate on gain for straw tube detector}\\
\vspace{0.2cm}

{\bf S.~Roy$^1$,
N.~Nandi$^2$,
R.~P.~Adak$^1$,
S. Biswas$^{1*}$,
S.~Das$^1$,
S.~K.~Ghosh$^1$,
S.~K.~Prasad$^1$,
S.~Raha$^1$\\}

\vspace{0.2cm}

  $^1$Bose Institute, Department of Physics and Centre for Astroparticle Physics and Space Science
(CAPSS), EN-80, Sector V, Kolkata-700091, India\\
$^2$Raja Peary Mohan College, 1 Acharya Dhruba Pal Road, Uttarpara, Hooghly, West Bengal- 712258, India\\
{\bf $^*$E-mail: saikat@jcbose.ac.in, saikat.ino@gmail.com, saikat.biswas@cern.ch}
\end{center}

\vspace{0.2cm}

\abstract{Basic R $\&$ D have been carried out with one small straw tube detector prototype with premixed gas of Ar+CO$_2$ in 70:30 and 90:10 ratio. The gain and the energy resolution are measured with Fe$^{55}$ X-ray source. Effect of temperature and pressure on these parameters are measured. The variation of gain and energy resolution with rate per unit length are also measured. The details of the test set-up, the method of measurement and the test results are presented in this paper.}

\vspace{0.2cm}


\section{Introduction}\label{intro}


The Compressed Baryonic Matter (CBM) experiment \cite{CBM} at the future Facility for Antiproton and Ion Research (FAIR) \cite{FAIR} in Darmstadt, Germany is designed to explore the QCD phase diagram in the region of moderate baryon densities. With CBM we will enter a new era of nuclear matter research by measuring rare diagnostic probes never observed before at FAIR energies, and thus CBM has a unique discovery potential. This will only be possible with the application of advanced instrumentation, including highly segmented and fast gaseous detectors. Keeping in mind the high interaction rate of FAIR, the Muon Chamber (MuCh) detector in CBM will use Gas Electron Multiplier (GEM) in the first two stations \cite{SB12, SB13, SB15, SB16, RPA16, RA}. Given the interaction rate of 10 MHz the expected particle flux on the first station will be about 3.1~MHz/cm$^2$. Maximum particle flux on the 3$^{rd}$ and 4$^{th}$ stations of the MuCh have been estimated to be 10 kHz/cm$^2$ and 3 kHz/cm$^2$, respectively, for central Au-Au collisions at 8 AGeV \cite{ZA16}. We are exploring the possibility of using straw tubes for the 3$^{rd}$ and 4$^{th}$ stations of CBM-MuCh \cite{CA07}.

Straw tubes are currently being used in large High Energy Physics (HEP) experiments as tracking detector with low material budget \cite{TA00}. A straw tube detector is basically a gas filled single channel drift tube with a conductive inner layer as cathode and a wire stretched along the axis as anode. When high voltage is applied between the wire and the tube an electric field is generated in the gas filled region. The electric field separates electrons and positive ions produced by an incident charged particle along its trajectory through the gas volume. The wire is kept at positive voltage and collects the electrons while the ions drift towards
the cathode. By choosing thin wires, with a diameter of a few tens of $\mu$m, the electric field
strength near the wire is made high enough to create an avalanche of electrons. Depending on the high voltage and the gas composition a gain of about 10$^{4}$ - 10$^{5}$ can be achieved \cite{panda}. The specific energy loss (dE/dx) of a charged particle in the straw gas volume can be used to identify the particle species and can be derived from the number of ionisation electrons per track length (dx) for the generated straw signal. Main advantage of using straw tube in a tracking system is reduction of material budget.

A systematic study of the characteristics of straw tube detectors with conventional Argon based gas mixtures has been performed. The motivation of this work is to study the rate handling capacity of straw tube detector and to measure the variation of gain and energy resolution using Fe$^{55}$ X-ray spectrum. The details of the measurement process and the experimental results are presented in this article.

\section{Experimental set-up}\label{setup}

\begin{figure}[htb!]
\begin{center}
\includegraphics[scale=0.28]{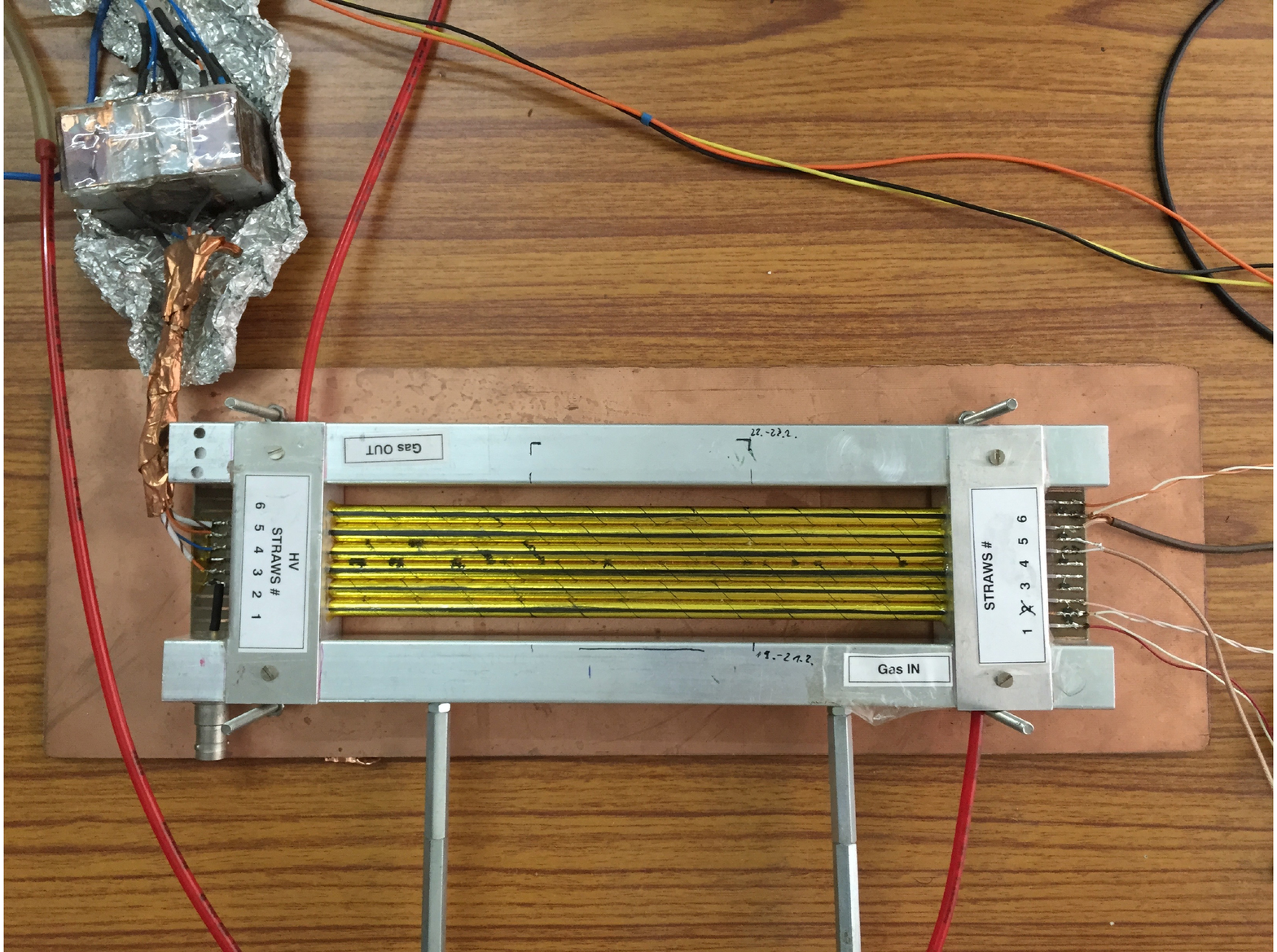}
\caption{The straw tube prototype : 6 straws, each of diameter 6 mm and length 20 cm.}
\label{straw}
 \end{center}
\end{figure}

\begin{figure}[htb!]
\begin{center}
\includegraphics[scale=0.6]{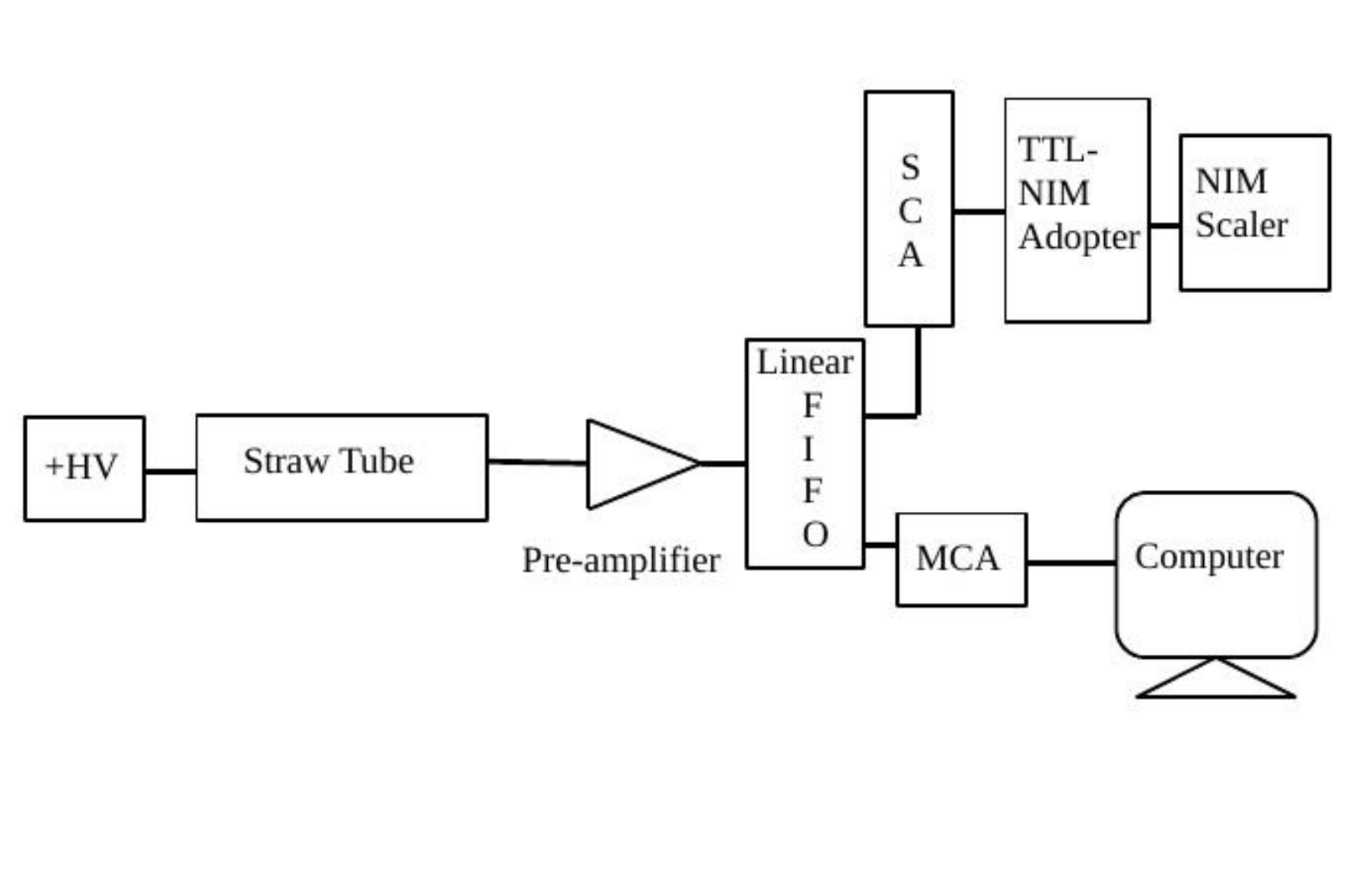}
\caption{Schematic representation of the electronics setup.}
\label{block}
 \end{center}
\end{figure}

A straw tube prototype shown in Figure \ref{straw} is obtained from JINR, Dubna,
Russia with 6 straws of diameter 6 mm and length 20 cm. Pre-mixed Ar/CO$_2$ in 70/30 and 90/10 volume ratio have been used for the different measurements. A constant gas flow rate of 3~l/h is maintained using a V{\"o}gtlin gas flow meter. The detector is tested using conventional NIM electronics. The positive high voltage (HV) is applied to one end of the central wire of the straws using a HV filter box and the signal is collected from the other end through a capacitor using LEMO connector. Single HV channel is used for each straw tube. The output signal from the straw is fed to a charge sensitive preamplifier (VV50-2) \cite{Preamp}. The gain of the preamplifier is 2~mV/fC. The output of the preamplifier is then fed to a linear Fan-in-Fan-out (linear FIFO) module. To measure the rate of incident particle the analog signal from the linear FIFO is put to a timing SCA (Single Channel Analyzer). The SCA is operated in integral mode and the lower level in the SCA is used as the threshold to the signal. The threshold is set at 1.3~V to reject all the noise. The discriminated TTL signal is fed to a TTL-NIM adopter and the output is counted using a NIM scaler. The count rate (i.e. counts per second) of the detector is then calculated. To obtain the energy spectrum, one output of the linear FIFO is fed to a Multi Channel Analyser (MCA). A schematic representation of the set-up is shown in Figure \ref{block}.

\section{Experimental results}\label{res}

\begin{figure}[htb!]
\begin{center}
\includegraphics[scale=0.45]{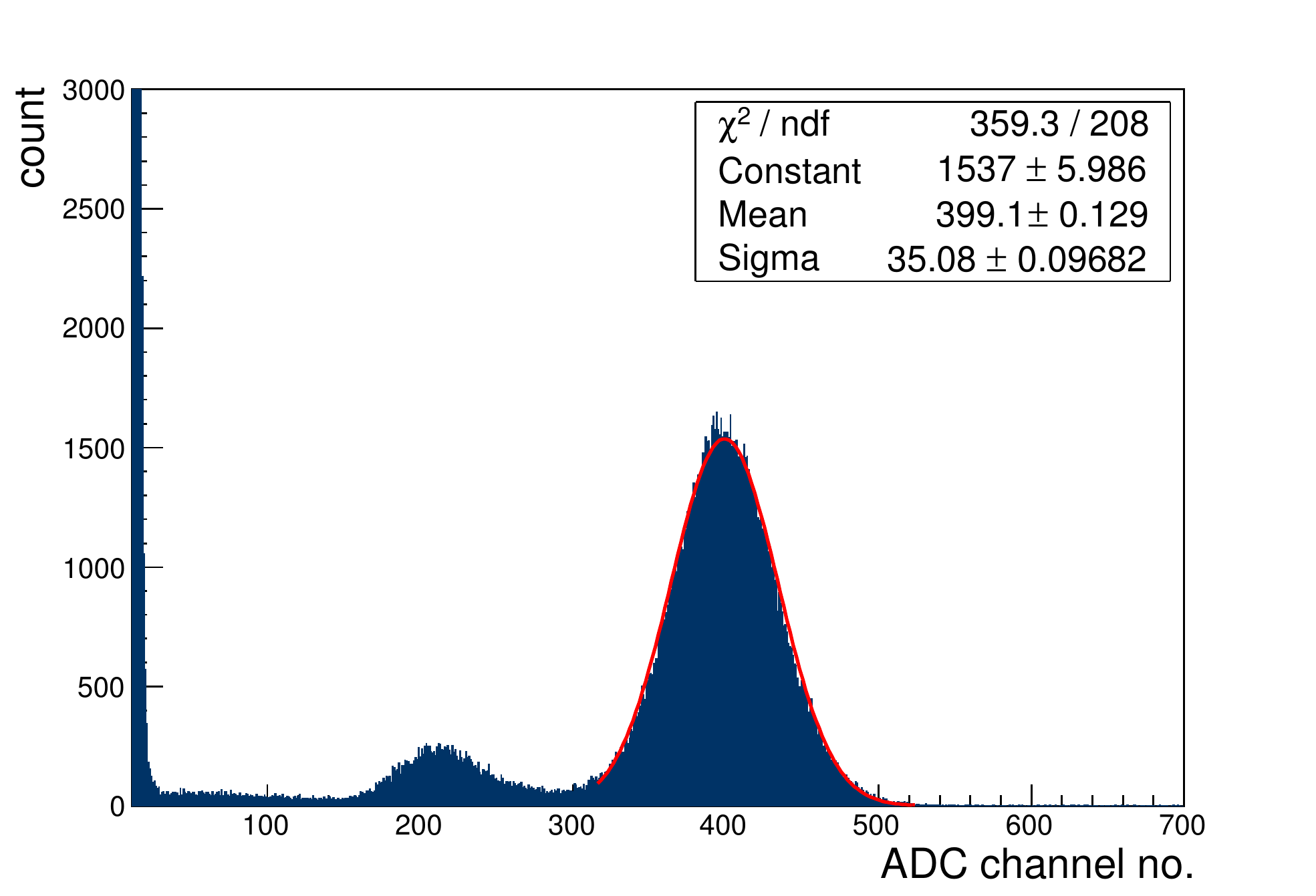}
\caption{Energy spectrum of the straw tube detector. The red line is the Gaussian fitting curve to the 5.9 keV peak.}
\label{spectrum}
 \end{center}
\end{figure}

The energy spectrum for the Fe$^{55}$ X-rays is obtained and the absolute gain and energy resolution are measured in the particular study. Figure \ref{spectrum} shows a typical spectrum recorded with a straw tube detector for Fe$^{55}$ source at a biasing voltage of 1650~V with Ar/CO$_2$ in 70/30 gas mixture. In this spectrum, the 
main peak (5.9 keV full energy peak) and the escape peak are clearly visible and well separated from the noise peak.
The gain of the detector has been measured by obtaining the mean position of 5.9 keV peak of Fe$^{55}$ X-ray spectrum with Gaussian fitting.

The expression for gain is given by:
\begin{eqnarray}
 gain &=& \frac{output~charge}{input~charge} \\
 &=& \frac{(mean~pulse~height/2 mV)~fC~\times~10^{15}}{No.~of~primary~electrons~\times~e~C} 
 \label{gaineq}
\end{eqnarray}
where the mean~pulse~height~for~5.9~keV~peak in ADC channel number is obtained by Gaussian fitting and that in mV is obtained from the ADC calibration curve (ADC channel no. vs pulse height). The preamplifier used in the set-up offers a gain of 2 mV/fC which has been used in the 
expression for gain. The input charge is the primary number of electrons 
produced in the gas detector as a result of total absorption 
of an X-ray photon of energy 5.9~keV, multiplied by the electronic charge ($e$). For each 5.9~keV Fe$^{55}$ X-ray photon exposed in Ar/CO$_2$ gas with 70/30 and 90/10 ratio, the 
number of primary electrons approximately produced are 212 and 222 respectively. 

\begin{figure}[htb!]
\begin{center}
\includegraphics[scale=0.45]{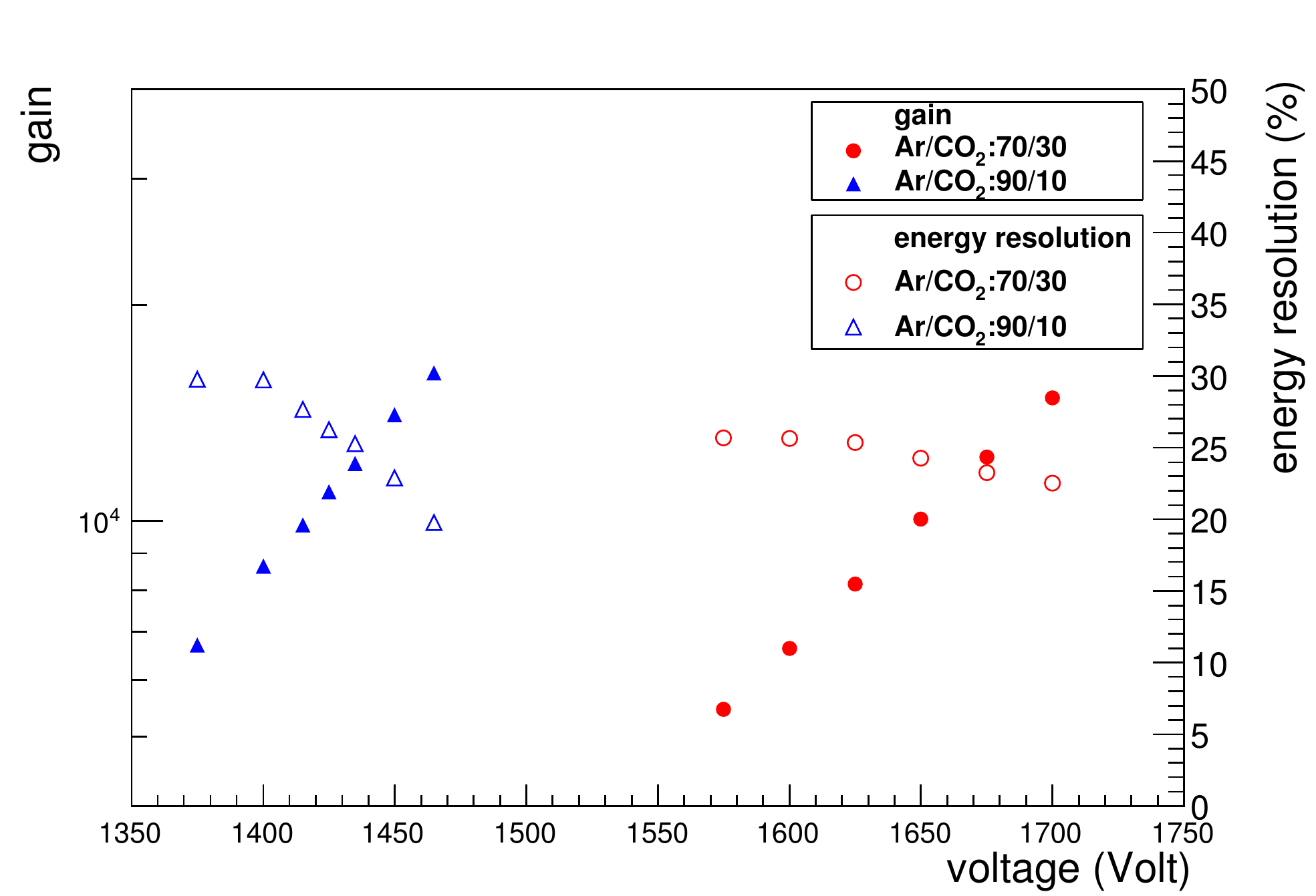}
\caption{The Gain and the energy resolution as a function of the voltage for both Ar/CO$_{2}$ 70/30 and 90/10 mixtures. The error bars are smaller than the symbols.}
\label{gainplot}
 \end{center}
\end{figure}

The energy resolution of the detector is defined as:
\begin{equation}
 \%~energy~resolution = \frac{sigma~\times~2.355}{mean} \times 100 \%  
\end{equation}
where the sigma and the mean are obtained from the Gaussian fitting of the 
spectrum. It is understood that a lower value of energy resolution means a 
better energy resolution. The gain and energy resolution have been measured, increasing the biasing voltage of the straw tube detector. It is observed that the 
gain increases exponentially whereas the energy resolution value 
decreases with the voltage as shown in Figure \ref{gainplot}.

It is well known that the gain of any gaseous detector depends significantly on the ratio of temperature and pressure, (T/p). The dependence of the gain (G) of a gaseous detector on absolute temperature and pressure is given by the relation \cite{MCA}
\begin{equation}
G(T/p) = Ae^{(B\frac{T}{p})}
 \label{gaintbypeq}
\end{equation}
where the parameters A and B are to be determined from the correlation plot.

The variation of the gain as a function of temperature and pressure is also studied for the straw tube detector from the energy spectrum obtained using the same Fe$^{55}$ source with Ar/CO$_2$ gas in 70/30 ratio. The detector is biased with 1650~V and is exposed to X-rays from the Fe$^{55}$ source and the energy spectra are recorded. Simultaneously the temperature (t in $^\circ C$), pressure (p in mbar) and relative humidity (RH in $\%$) are also recorded using a data logger, built in-house \cite{Sahu}. The measurement is done for a time period of $\sim$~340 minutes.

\begin{figure}[htb!]
\begin{center}
\includegraphics[scale=0.45]{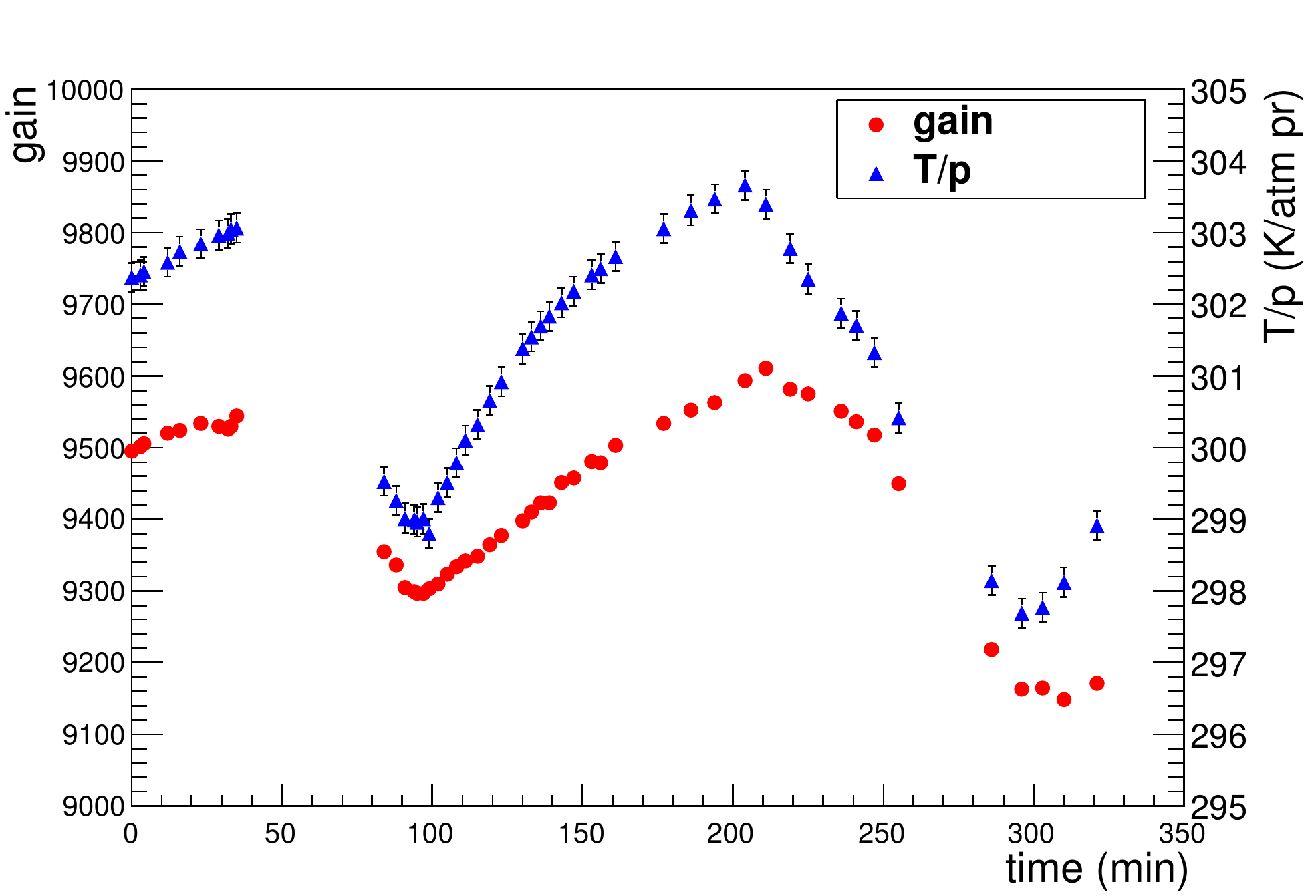}
\caption{Variation of the measured gain and T/p as a function of the time. The error bars are smaller than the symbols for gain.}\label{gaintime}
\end{center}
\end{figure}


The absolute gain of the detector is calculated from the formula given in equation~\ref{gaineq}. The variation of the measured gain is plotted as a function of period of operation in Figure~$\ref{gaintime}$. The variation of the T/p as a function of the total period of operation is also plotted in Figure~$\ref{gaintime}$, where T (= t+273) is the absolute temperature in Kelvin and p (p in mbar/1013) is in the unit of atmospheric pressure. 

\begin{figure}[htb!]
\begin{center}
\includegraphics[scale=0.45]{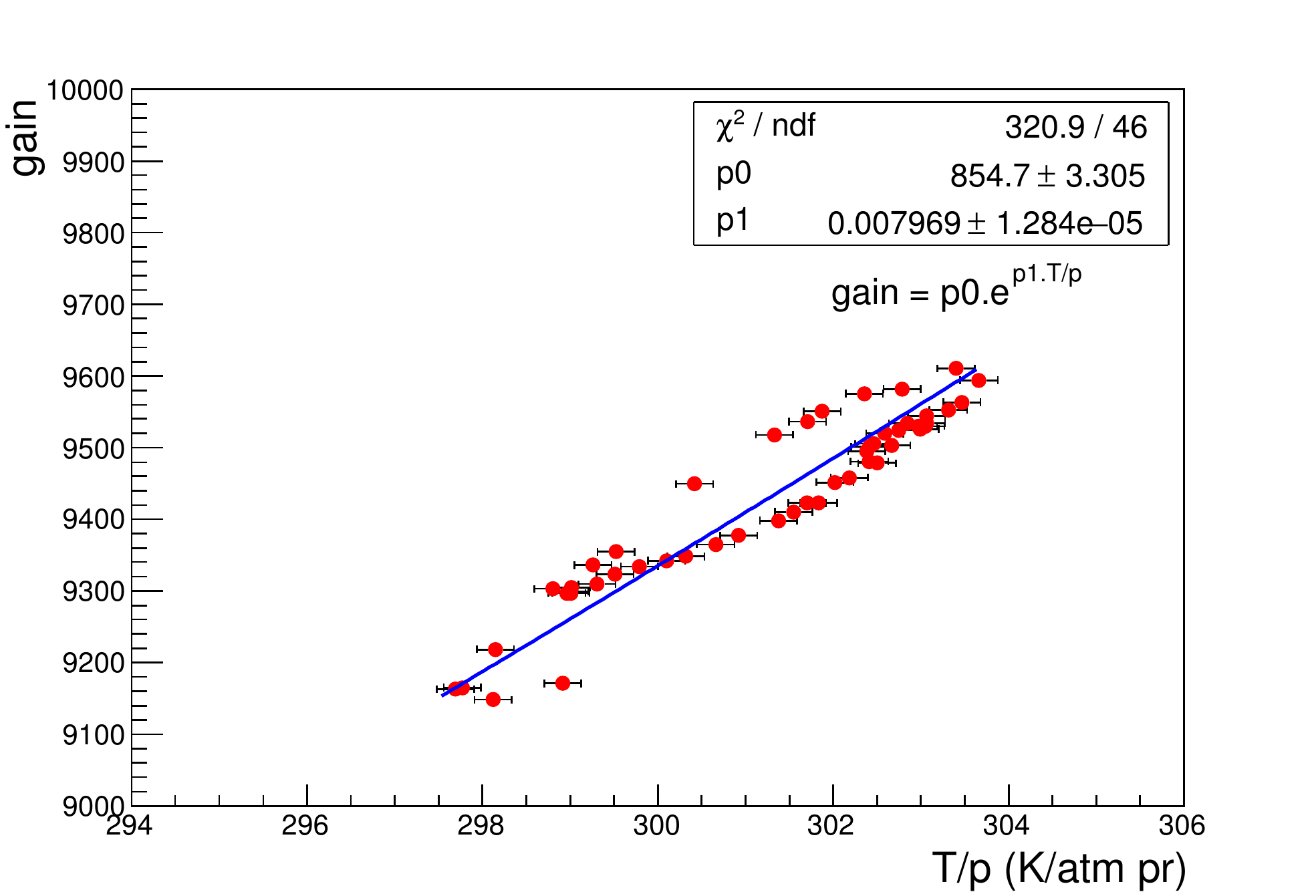}
\caption{Correlation plot: Variation of the gain as a function of T/p.}\label{gainTbyp}
\end{center}
\end{figure}

The correlation plot, i.e. the gain is plotted as a function of T/p and fitted with the function given by equation~\ref{gaintbypeq} and is shown in Figure~$\ref{gainTbyp}$ (in Figure~$\ref{gainTbyp}$ the parameters A and B are marked as p0 and p1 respectively).


The values of the fit parameters A and B obtained, are 854.7~$\pm$~3.305 and 0.0079~$\pm$~1.284~$\times$~10$^{-5}$ atm pr/K. Using the fit parameters, the gain is normalised by using the relation:

\begin{equation}
gain_{normalised} = \frac{gain_{measured}}{Ae^{(B\frac{T}{p})}}
\end{equation}

\begin{figure}[htb!]
\begin{center}
\includegraphics[scale=0.45]{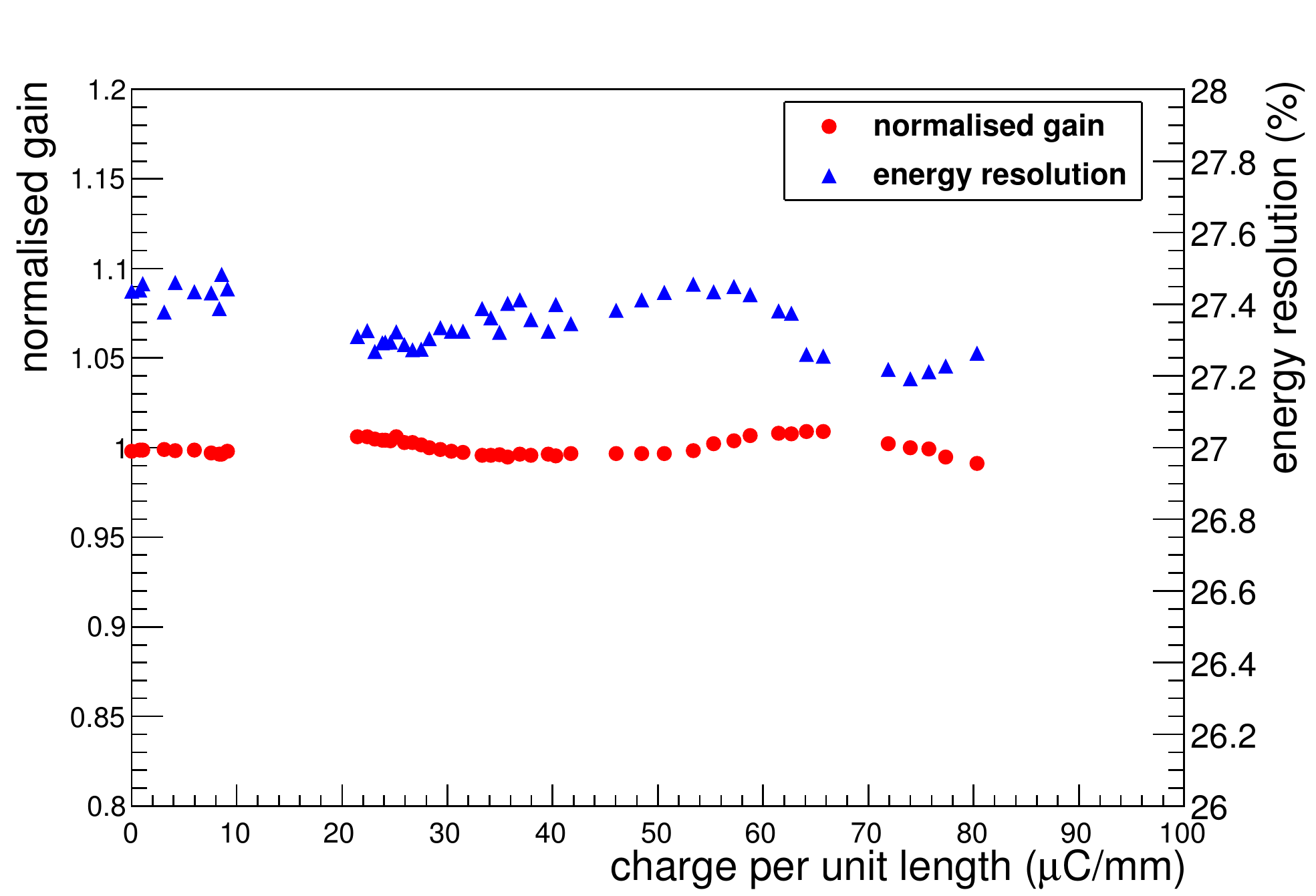}
\caption{Variation of the normalised gain and energy resolution as a function of the charge per unit length i.e. dQ/dL. The error bars are smaller than the symbols.}\label{normvscharge}
\end{center}
\end{figure}

\begin{figure}[htb!]
\begin{center}
\includegraphics[scale=0.45]{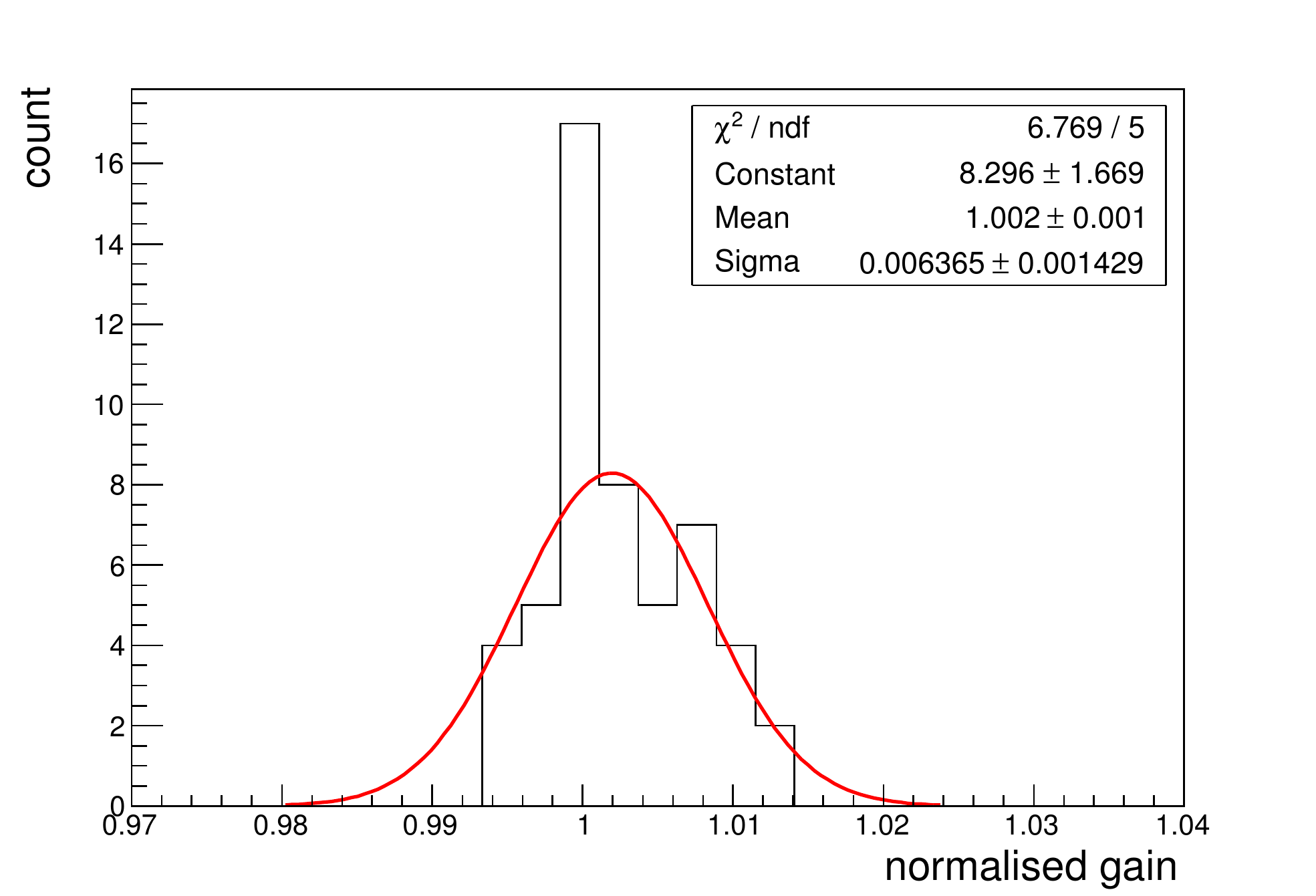}
\caption{The distribution of the normalised gain fitted with a Gaussian function.}\label{hist}
\end{center}
\end{figure}

To check the stability of the detector, the normalised gain is plotted against the total charge accumulated per unit irradiated length of the detector which is directly proportional to time. The charge accumulated at a particular time is calculated by
\begin{equation}
\frac{dq}{dL} = \frac{r \times n \times e \times G \times dt}{dL} 
\end{equation}
where, $r$ is the measured rate in Hz incident on a particular length of the detector, $dt$ is the time in second, $n$ is the number of primary electrons for a single X-ray photon, $e$ is the electronic charge, $G$ is the gain and $dL$ is the irradiated length. The normalised gain as a function of the charge accumulated per unit length is shown in Figure~$\ref{normvscharge}$. There is a fluctuation around 1 in the normalised gain value as shown in Figure~$\ref{normvscharge}$. The distribution of the normalised gain fitted with a Gaussian function is shown in Figure~$\ref{hist}$. The mean of the Gaussian distribution has been found to be around 1.002 with a sigma of 0.006 as shown in Figure~$\ref{hist}$. In this study an accumulation of charge per unit length $\sim$~80 $\mu$C/mm is achieved.

\begin{figure}[htb!]
\begin{center}
\includegraphics[scale=0.45]{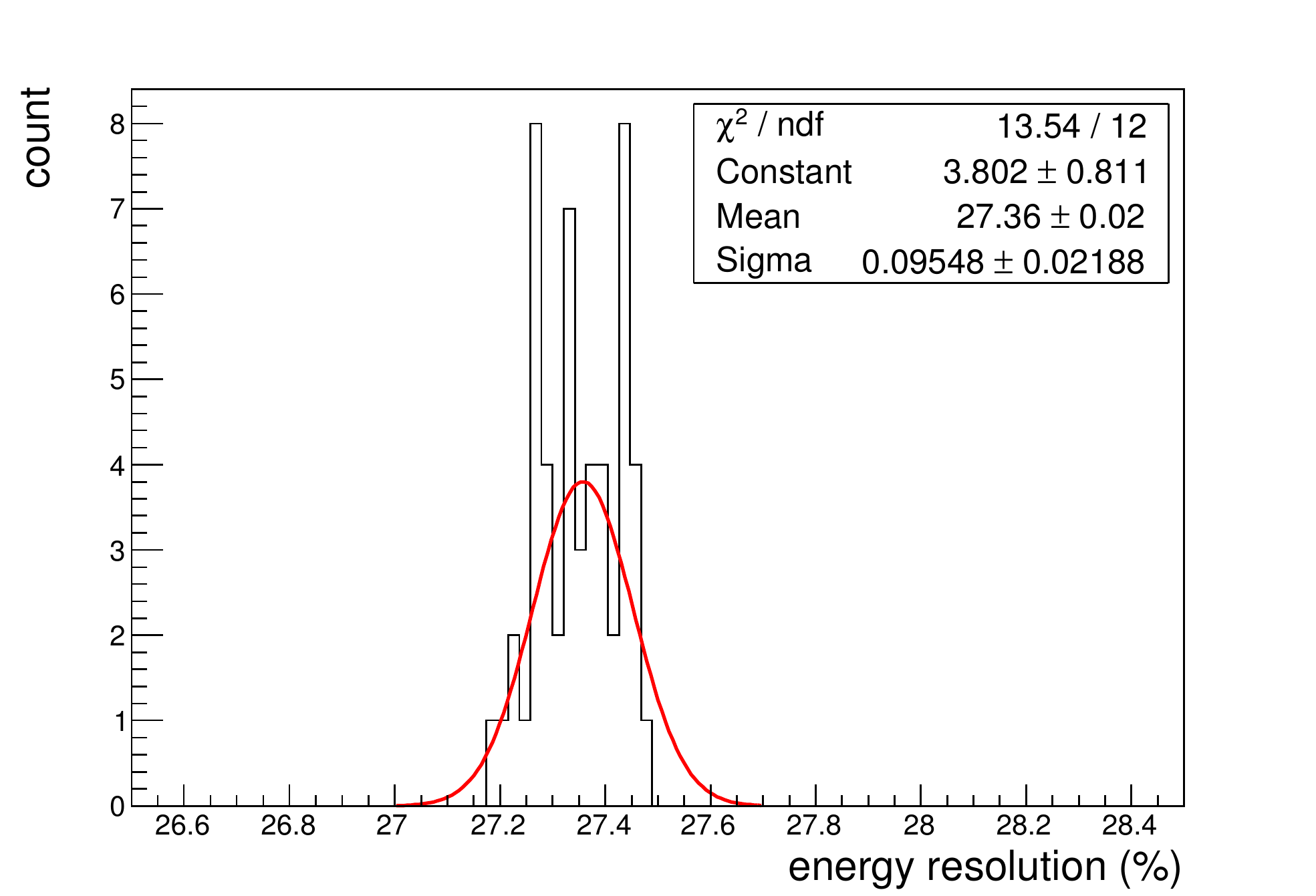}
\caption{The distribution of the energy resolution fitted with a Gaussian function.}\label{resohist}
\end{center}
\end{figure}
\begin{figure}[htb!]
\begin{center}
\includegraphics[scale=0.45]{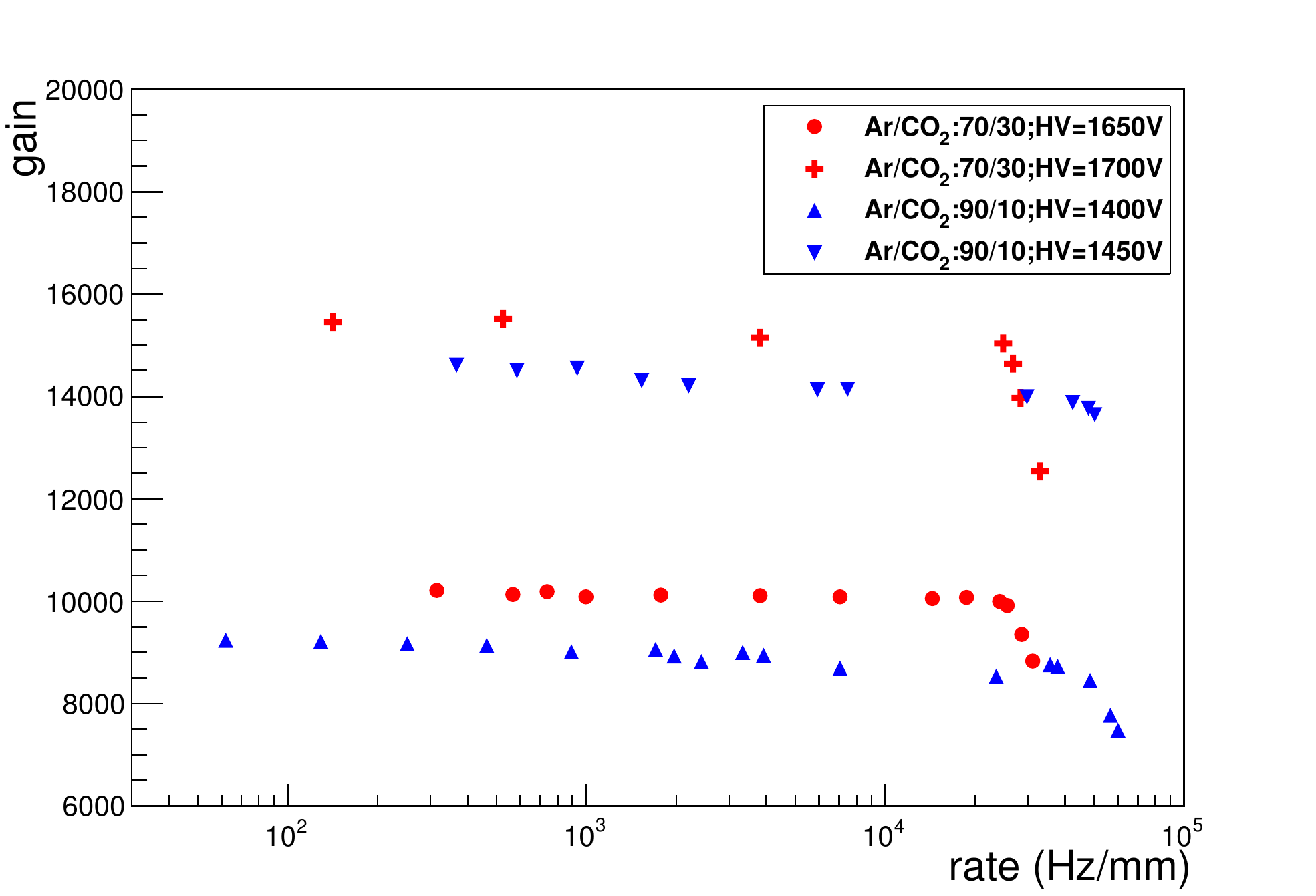}
\caption{Gain as a function of rate for both Ar/CO$_{2}$ 70/30 and 90/10 mixtures. The error bars are smaller than the symbols.}
\label{gainrateplot}
 \end{center}
\end{figure}
\begin{figure}[htb!]
\begin{center}
\includegraphics[scale=0.45]{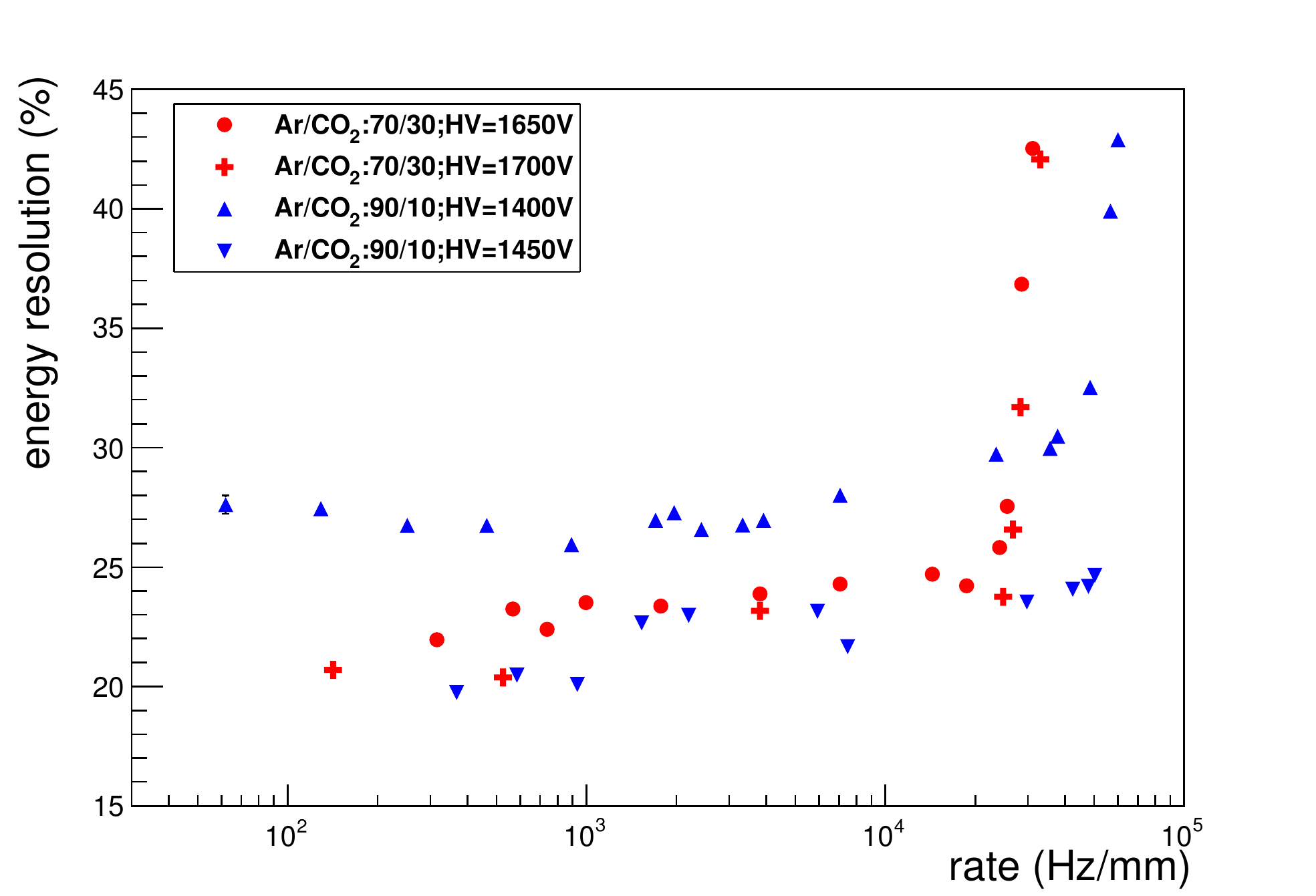}
\caption{Energy resolution as a function of rate for both Ar/CO$_{2}$ 70/30 and 90/10 mixtures. The error bars are smaller than the symbols.}
\label{resorateplot}
 \end{center}
\end{figure}

In this study the energy resolution is also measured for each spectrum. The energy resolution as a function of charge accumulated per unit length (or time, as charge accumulated per unit length is directly proportional to time) is also shown in Figure~$\ref{normvscharge}$. The distribution of energy resolution as shown in Figure~$\ref{resohist}$ shows that during this time of $\sim$~340 minutes the mean energy resolution is 27.36\% with a sigma of 0.09. 

The variation of the gain and the energy resolution of the straw tube detector is measured varying the rate of incident X-ray photons on the detector. A collimator made with perspex is used for the Fe$^{55}$ source to change the rate of emitted X-ray. The collimator opening is changed to vary the rate of particles incident on the detector. The energy spectrum is obtained for each setting of the collimator. These measurements are performed with both Ar/CO$_{2}$ 70/30 and 90/10 gas mixtures. For Ar/CO$_{2}$ 70/30 the measurements are performed keeping the HV to the straws at 1650~V and 1700~V whereas for Ar/CO$_{2}$ 90/10 it is done for HV 1400~V and 1450~V. For Ar/CO$_{2}$ 70/30 the gain and energy resolution are measured from a rate of about 200~Hz/mm to about 3~$\times$~10$^4$~Hz/mm and that for Ar/CO$_{2}$ 90/10 are performed for about 100~Hz/mm to about 6~$\times$~10$^4$~Hz/mm. Measured gain and energy resolution as a function of X-ray rate per unit length are shown in Figure~\ref{gainrateplot} and Figure~\ref{resorateplot} respectively. It is observed that for Ar/CO$_{2}$ 70/30 the gain and the energy resolution remains constant up to a rate of about 2~$\times$~10$^4$~Hz/mm then the gain decreases and energy resolution value increases with rate. Similar effect is observed for Ar/CO$_{2}$ 90/10 as well, where the gain and energy resolution remains constant up to a rate of about 3.2~$\times$~10$^4$~Hz/mm.

For higher rates the gain ($G$) is fitted with a function \cite{Sauli}
\begin{equation}
G = Pe^{- Q~.~R }
\end{equation}
where $P$ and $Q$ are the fit parameters and $r$ is the rate.

For higher rates the energy resolution is fitted with a function
\begin{equation}
energy~resolution = P'e^{Q'~.~R }
\end{equation}
where $P'$ and $Q'$ are the fit parameters and $R$ is the rate.

The numerical values of $P$, $Q$, $P'$ and $Q'$ are tabulated in Table~\ref{table1}. 

\begin{table}[h!]
\caption{\label{table1}Values of the fit parameters.}
\begin{center}

\begin{tabular}{|c|c|c|c|c|c|} \hline
Gas & Voltage   &  $P$  & $Q$  & $P'$  & $Q'$   \\
mixture & (Volt) &  &  &  &     \\
Ar/CO$_{2}$ &  &  &  &  &     \\ \hline
70/30 & 1650  &  1.58 & 1.86 & 4.72 & 7.07  \\
 &  &  $\times$10$^4$ & $\times$10$^{-5}$ &  & $\times$10$^{-5}$  \\ \hline
 70/30 & 1700  &  2.75 & 2.38 & 4.25 & 6.98  \\
 &  &  $\times$10$^4$ & $\times$10$^{-5}$ &  & $\times$10$^{-5}$  \\ \hline
 90/10 & 1400  &  1.08 & 5.67 & 24.77 & 6.70  \\
 &  &  $\times$10$^4$ & $\times$10$^{-6}$ &  & $\times$10$^{-6}$  \\ \hline
 90/10 & 1450  & 1.53 & 2.26 & 21.41 & 2.71  \\
 &  &  $\times$10$^4$ & $\times$10$^{-6}$ &  & $\times$10$^{-6}$  \\ \hline

\end{tabular}\\
 \end{center}

\end{table}











It is observed that for Ar/CO$_2$ 70/30 gas mixture the detector is operated at relatively higher voltages and in this case decrease of gain with rate started at relatively lower rate.

\section{Conclusions and outlook}
A systematic study on the basic characteristics of straw tube detector is performed using conventional NIM electronics. In this study Ar/CO$_2$ gas mixture is used both in 70/30 and 90/10 volume ratio. The gain and energy resolution are measured from the energy spectrum obtained using Fe$^{55}$ X-ray source. To check the effect of temperature and pressure on the gain and energy resolution a continuous measurement is performed. Same Fe$^{55}$ X-ray source is used to irradiate the detector and to obtain the spectrum. The measured gain is normalised by T/p corrected gain. The normalised gain is found to be stable with an average value of 1.002 with a sigma of 0.006 for a duration of $\sim$~340 minutes which is equivalent to an accumulation of charge per unit length $\sim$~80 $\mu$C/mm. In this study main emphasis is given on the variation of gain and energy resolution of the straw tube detector with X-ray rate. The gain and the energy resolution remain constant up to a rate of about about 2~$\times$~10$^4$~Hz/mm and 3.2~$\times$~10$^4$~Hz/mm for Ar/CO$_2$ 70/30 and 90/10 respectively. Beyond these quoted values gain decreases and the energy resolution increases with the increase of rate. Keeping in mind the particle flux straw tube is an option to be used in the 3$^{rd}$ and 4$^{th}$ stations of CBM-MuCh. Possibility to use the straw tube detector in CBM MuCh is under investigation.

\section{Acknowledgements}
We would like to thank Late Prof.~Vladimir~Peshekhonov of JINR, Dubna for providing the straw tube prototype and Dr. Subhasis Chattopadhyay, Mr. J. Saini of VECC, Kolkata, Dr.~Christian~J.~Schmidt of 
GSI Detector Laboratory for valuable discussions in the course of the study. This work is partially supported by the research grant of CBM-MUCH project from BI-IFCC, Department of Science and Technology, Govt. of India. This work is also partially supported by the research grant SR/MF/PS-01/2014-BI from Department of Science and Technology, Govt. of India. S. Biswas acknowledges the support of DST-SERB Ramanujan Fellowship (D.O. No. SR/S2/RJN-02/2012). R.~P.~Adak acknowledges the support UGC order no - 20-12/2009 (ii) EU-IV. S. Biswas would like to thank Ms. S. Rudra for a valuable discussion on this article.


\noindent
\begin{thebibliography}{50}


\bibitem{CBM} http://www.fair-center.eu/for-users/experiments/cbm.html .

\bibitem{FAIR} http://www.fair-center.eu/ .

\bibitem{SB12} S.~Biswas et al., 2013 JINST 8 C12002 doi:10.1088/1748-0221/8/12/C12002.

\bibitem{SB13} S. Biswas et al., Nucl. Instrum. Meth. A 718 (2013) 403.

\bibitem{SB15} S. Biswas et al., Nucl. Instrum. Meth. A 800 (2015) 93.

\bibitem{SB16} S. Biswas et al., Nucl. Instrum. Meth. A 824 (2016) 504.

\bibitem{RPA16} R. P. Adak et al., Nucl. Instrum. Meth. A 846 (2017) 29. [arXiv:1604.02899v2].

\bibitem{RA} R.P. Adak et al., 2016 JINST 11 T10001 doi:10.1088/1748-0221/11/10/T10001.

\bibitem{ZA16} Z. Ahammed et al., CBM Progress Report (2016) 78.

\bibitem{CA07} C. Adorisio et al., Nucl. Instrum. Meth. A 575 (2007) 532.

\bibitem{TA00} T. Akesson et al., Nucl. Instrum. Meth. A 449 (2000) 446.  

\bibitem{panda} PANDA Collaboration, Technical Design Report for the: PANDA Straw Tube Tracker, arXiv:1205.5441.

\bibitem{Preamp} CDT CASCADE Detector Technologies GmbH, Hans-Bunte-Str. 8-10, 69123 Heidelberg, Germany, www.n-cdt.com.

\bibitem{MCA} M.C. Altunbas et al., Nucl. Instrum. Meth. A 515 (2003) 249.

\bibitem{Sahu} S. Sahu et al., 2017 JINST 12 C05006 doi:10.1088/1748-0221/12/05/C05006.

\bibitem{Sauli} F. Sauli, Principles of operation of multi wire proportional and drift chambers, CERN 77-09, 3 May 1977.





\end{thebibliography}
\end{document}